\documentclass[aps,prd,groupedaddress,nofootinbib,showpacs]{revtex4}
\usepackage{graphicx,amssymb}

\begin{document}
\title{Fourth order gravity and experimental constraints on Eddington parameters}
\author{S. Capozziello$^1$, A. Stabile$^2$, A. Troisi$^1$}

\affiliation{$^1$Dipartimento di Scienze Fisiche, Univ. di Napoli "Federico II", and INFN, Sez. di Napoli,
Compl. Univ. di Monte S. Angelo, Ed. N, via Cinthia, 80121 - Napoli, Italy \\
$^2$Dipartimento di Fisica "E.R. Caianiello",  Univ. di Salerno, and INFN, Sez. di Napoli, Gruppo Coll. di
Salerno, via S. Allende, 84081 - Baronissi (Salerno), Italy}
\date{\today}

\begin{abstract}
PPN-limit of alternative theories of gravity represents a still
controversial matter of debate and no definitive answer has been
provided, up to now, about this issue.
 By using the definition of the
PPN-parameters $\gamma$ and $\beta$ in term of  $f(R)$ theories of
gravity, we show that a family of third-order polynomial theories,
in the Ricci scalar $R$,  turns out to be compatible with the
PPN-limit and the deviation from General Relativity, theoretically
predicted, can agree with experimental data.
\end{abstract}


\maketitle


\vspace{5.mm}

%\section{Introduction}

\noindent 1. General Relativity (GR) is the cornerstone theory
among the several attempts proposed to describe gravity. It
represents an elegant approach furnishing several phenomenological
predictions and its validity, in the Newtonian limit regime, is
experimentally probed  up to the Solar System scales. However,
also at these scales, some conundrums come out as the indications
of an apparent, anomalous, long--range acceleration revealed from
the data analysis of Pioneer 10/11, Galileo, and Ulysses
spacecrafts which are difficult to be framed in the standard
scheme of GR and its low energy limit
\cite{anderson_1,anderson_2}. Furthermore, at  galactic distances,
huge bulks of dark matter are needed to provide realistic models
matching with observations. In this case, retaining GR and its low
energy limit  implies the introduction of an actually unknown
ingredient. We face a similar situation even at larger scales:
clusters of galaxies are gravitationally stable and bound  only if
large amounts of dark matter are supposed in their potential
wells. Finally, an unknown form of dark energy is required to
explain the observed accelerated expansion of cosmic fluid.
Summarizing, almost $95\%$ of matter-energy content of the
universe is unknown in the framework of Standard Cosmological
Model while we can experimentally probe only gravity and ordinary
(baryonic  and radiation)  matter. Considering another point of
view, anomalous acceleration (Solar System), dark matter (galaxies
and galaxy clusters), dark energy (cosmology) could be nothing
else but the indications that shortcomings are present in GR and
gravity is an interaction depending on the scale. The assumption
  of a linear Lagrangian density in the Ricci scalar $R$ for the Hilbert-Einstein action
could be too simple to describe gravity at any scale and more
general approaches should be pursued to match observations. Among
these schemes, several motivations suggest to generalize GR by
considering gravitational actions where generic functions of
curvature invariants are present. Specifically, actions of the
form
\begin{equation}
\label{f(R)} \mathcal{A}=\int
d^4x\sqrt{-g}\biggl[f(R)+\mathcal{L}_m\biggr]\,,
\end{equation}
where $f(R)$ is an analytic function of $R$ and $\mathcal{L}_m$ is
the standard matter Lagrangian density, result particularly
interesting. The variation of (\ref{f(R)}) gives rise to
fourth-order field equations
\begin{equation}\label{5}
G_{\alpha \beta} = R_{\alpha\beta}-\frac{1}{2}g_{\alpha\beta}R =
T^{curv}_{\alpha\beta}+T^{m}_{\alpha\beta}/f^\prime(R)
\end{equation}
where the {\it curvature stress\,-\,energy tensor} is defined as
\begin{equation} \label{6}
T^{curv}_{\alpha\beta}\,=\,\frac{1}{f'(R)}\left\{\frac{1}{2}g_{\alpha\beta}\left[f(R)-Rf'(R)\right]
+f'(R)^{;\mu\nu}(g_{\alpha\mu}g_{\beta\nu}-g_{\alpha\beta}g_{\mu\nu})
\right\}\,,
\end{equation}
 with  prime denoting derivative with respect
to $R$ and $T^{m}_{\alpha\beta}$  the standard matter
contribution. For $f(R) = R$, the curvature stress\,-\,energy
tensor identically vanishes and Eqs.(\ref{5}) reduce to the
standard second\,-\,order Einstein field equations. From
Eq.(\ref{5}), it is clear that the curvature stress\,-\,energy
tensor  plays the role of a further source term in the field
equations so that it can be considered as an effective fluid of
purely geometric origin.

This approach is physically motivated by several unification
theories of fundamental interactions and by the field quantization
on curved space-times \cite{birrell}. At cosmological level,  it
is well known that further curvature contributions can solve
shortcomings at early epochs (giving rise to inflationary
solutions \cite{starobinsky}) and explaining the today observed
accelerated behavior by a sort of {\it curvature quintessence},
i.e. curvature can act as a fluid implementing acceleration
\cite{curv-quint,noi-ijmpd,carroll,odintsov_1,odintsov_2,odintsov_3}.
This result can be achieved in metric and affine (Palatini)
approaches
\cite{noi-review,francaviglia_1,francaviglia_2,palatini_1,palatini_2,palatini_3,palatini_4,palatini_5,palatini_6,
multamaki}. In addition, reversing the problem, one can
reconstruct the form of the gravity Lagrangian by observational
data of cosmological relevance through a "back scattering"
procedure. All these facts suggest that the function $f(R)$ should
be more general than the linear Hilbert-Einstein one implying that
higher order gravity could be a suitable approach to solve GR
shortcomings without introducing mysterious ingredients as dark
energy and dark matter (see e.g. \cite{mimicking,newtlim}.

In recent papers, some authors  have confronted this kind of
theories even with the Post Parameterized Newtonian prescriptions
in metric and Palatini approaches. However, the results seem
controversial since in some cases \cite{olmo_1,olmo_2} it is
argued that GR is always valid at Solar System scales and there is
no room for other theories; nevertheless, some other studies
\cite{ppn-noi, allemandi-ruggiero} find that recent experiments as
Cassini and Lunar Laser Ranging allow the possibility that
extended theories of gravity could be seriously taken into
account. In particular, it is possible to define PPN-parameters in
term of $f(R)$ functions and several classes of fourth order
theories result compatible with experiments in Solar System
\cite{ppn-noi}.

In this letter, we follow a different approach. Starting  from the
definitions of  PPN-parameters in term of a generic analytic
 function $f(R)$ and its derivatives,  we  deduce a class of fourth
order theories,  compatible with data, by means of an inverse
procedure which allows to compare PPN-conditions   with data. As a
matter of fact, it is possible to show that a third order
polynomial, in the Ricci scalar, is compatible with observational
constraints on PPN-parameters. The degree of deviation from GR
depends on the experimental estimate of PPN-parameters.

%\section{PPN-parameters in Scalar-Tensor and Fourth-Order Gravity}
%\noindent 2.

A useful method to take into account deviation with respect to GR
is to develop expansions about the GR solutions up to some
perturbation orders. A standard approach is the
Parameterized-Post-Newtonian (PPN) expansion of the Schwarzschild
metric. In isotropic coordinates, it is
\begin{equation}\label{schwarz-eddington}
ds^2=\biggl(\frac{1-\frac{r_g}{4\tilde{r}}}{1+\frac{r_g}{4\tilde{r}}}\biggr)^2dt^2-\biggl(1+\frac{r_g}{4\tilde{r}}\biggr)^{4}
(d\tilde{r}^2+\tilde{r}^2d\Omega^2)
\end{equation}
where $r_g\,=\,2GM/c^2$ is the Schwarzschild radius.  Eddington
parameterized deviations with respect to GR, considering  a Taylor
series  in term of $r_g /\tilde{r}$ assuming that in Solar System,
the limit $r_g/\tilde{r}) \ll 1$ holds \cite{will}. The resulting
metric is
\begin{equation}\label{schwarz-eddington2}
ds^2\simeq\biggl[1-\alpha\frac{r_g}{\tilde{r}}+\frac{\beta}{2}\biggl(\frac{r_g}{\tilde{r}}\biggr)^2+\dots\biggr]dt^2
-\biggl(1+\gamma\frac{r_g}{\tilde{r}}+\dots\biggr)\biggl(d\tilde{r}^2+\tilde{r}^2
d\Omega^2\biggr)\end{equation} where $\alpha$, $\beta$ and
$\gamma$ are unknown dimensionless parameters (Eddington
parameters) which parameterize deviations with respect to GR. The
reason to carry out this expansion up to the order $(r_g
/\tilde{r})^2$ in $g_{00}$ and only to the order $(r_g/\tilde{r})$
in $g_{ij}$ is that, in applications to celestial mechanics,
$g_{ij}$  always  appears multiplied by an extra factor
$v^2\backsim (M/\tilde{r})$. It is evident that the standard GR
solution for a  spherically symmetric gravitational system in
vacuum is obtained for $\alpha = \beta = \gamma = 1$ giving again
the Schwarzschild solution. Actually, the parameter $\alpha$ can
be settled to the unity due to the mass definition  of the system
itself \cite{will}. As a consequence, the expanded metric
(\ref{schwarz-eddington2}) can be recast in the form\,:

\begin{equation}
ds^2
\simeq\biggl[1-\frac{r_g}{r}+\frac{\beta-\gamma}{2}\biggl(\frac{r_g}{r}\biggr)^2+\dots\biggr]dt^2
 - \biggl[1+\gamma\frac{r_g}{r}+\dots\biggr]dr^2-r^2d\Omega^2\,,
\end{equation}
where we have restored the standard spherical coordinates by means
of the transformation
$r\,=\,\tilde{r}\left(1+\frac{r_g}{4\tilde{r}}\right)^2$. The two
parameters $\beta,\ \gamma$ have a   physical interpretation. The
parameter $\gamma$ measures the amount of curvature of space
generated by a body of mass $M$ at radius $r$. In fact, the
spatial components of the Riemann curvature tensor are, at
post-Newtonian order,
\begin{equation}
R_{ijkl}=\frac{3}{2}\gamma\frac{r_g}{r^3}N_{ijkl}
\end{equation}
independently of the gauge choice, where $N_{ijkl}$ represents the
geometric tensor properties (e.g. symmetries of the Riemann tensor
and so on). On the other side, the parameter $\beta$ measures the
amount of non-linearity ($\sim (r_g/r)^2 $) in the $g_{00}$
component of the metric. However, this statement is valid only in
the standard post-Newtonian gauge.

If one takes into account a more general theory of gravity, the
calculation of the PPN-limit can be performed following a well
defined pipeline which straightforwardly generalizes the standard
GR case \cite{will}. A significant development in this sense has
been pursued by Damour and Esposito-Farese
\cite{damour1,damour_21,damour_22,damour_23} which have approached
to the calculation of the PPN-limit of scalar-tensor gravity by
means of a conformal transformation $\tilde{g}_{\mu\nu}\,=\,
F(\phi)g_{\mu\nu}$ to the standard Einstein frame. In fact, a
general scalar-tensor theory
\begin{equation}\label{scaten-L}
\mathcal{A}=\int
d^4x\sqrt{-g}\biggl[F(\phi)R+\frac{1}{2}\phi_{;\mu}\phi^{;\mu}-V(\phi)+\mathcal{L}_m\biggr]\,,
\end{equation}
where $V(\phi)$ is the self-interaction potential and $F(\phi)$
the non-minimal coupling, can be recast as\,:
\begin{equation}\label{scaten-L-conf}
\tilde{\mathcal{A}}=\int
d^4x\sqrt{-\tilde{g}}\biggl[-\frac{1}{2}\tilde{R}+\frac{1}{2}\varphi_{;\mu}\varphi^{;\mu}-\tilde{V}(\varphi)+
\tilde{\mathcal{L}}_m\biggr]
\end{equation}
where
\begin{equation}
\biggl(\frac{d\varphi}{d\phi}\biggr)^2=\frac{3}{4}\biggl(\frac{d \ln
F(\phi)}{d\phi}\biggr)^2+\frac{1}{2F(\phi)}\,,
\end{equation}
$8\pi G=1$, $\tilde{V}(\varphi)\,=\,F^{-2}(\phi)V(\phi)$,
$\tilde{\mathcal{L}}_m\,=\,F^{-2}(\phi)\mathcal{L}_m$. The first
consequence of such a transformation is that now the non-minimal
coupling is transferred to the ordinary-matter sector. In fact,
the Lagrangian $\tilde{\mathcal{L}}_m$ is dependent not only on
the conformally transformed metric $\tilde{g}_{\mu\nu}$ and matter
fields but it is even  characterized by the coupling function
$F(\phi)$ \cite{faraoni}. This scheme  provides several
interesting results up to obtain an intrinsic definition of
$\gamma,\ \beta$ in term of the non-minimal coupling function
$F(\phi)$. The analogy between scalar-tensor gravity and higher
order theories of gravity has been widely demonstrated
\cite{teyssandier83,schmidtNL,wands:cqg94}. Here an important
remark is in order. The analogy between scalar-tensor gravity and
fourth order gravity, although mathematically straightforward,
requires a careful physical analysis. Recasting fourth-order
gravity as a scalar-tensor theory, often the following  steps, in
terms of a generic scalar field $\psi$, are considered
\begin{equation}\label{o'hanlon} f(R)+\mathcal{L}_m \,\rightarrow\,
F'(\psi)R+F(\psi)-F'(\psi)\psi+\mathcal{L}_m \,\rightarrow\,
F'(\psi)R-V(\psi)+\mathcal{L}_m\,,
\end{equation}
where, by analogy, $\psi \rightarrow R$ and the "potential" is
$V(\psi)=F(\psi)-F'(\psi)\psi$. Clearly the kinetic term is not
present so that (\ref{o'hanlon}) is usually referred as a
Brans-Dicke description of $f(R)$ gravity where $\omega_{BD}=0$.
This is the so-called O'Hanlon Lagrangian \cite{teyssandier83}.
However, the typical Brans-Dicke action is
\begin{equation}
\mathcal{A}=\int d^4x\sqrt{-g}\left[\psi
R-\omega_{BD}\frac{\psi_{;\mu}\psi^{;\mu}}{\psi}+
\mathcal{L}_m\right]\,,
\end{equation}
where no scalar field potential is present and $\omega_{BD}$ is a
constant. In summary, O'Hanlon Lagrangian has a potential but has
no kinetic term, while Brans-Dicke Lagrangian has a kinetic term
without potential. The most general situation is in
(\ref{scaten-L}) where we have non-minimal coupling, kinetic term,
and scalar field potential. This means that fourth-order gravity
and scalar tensor gravity can be "compared" only by means of
conformal transformations where kinetic and potential terms are
preserved. In particular, it is misleading to state that PPN-limit
of fourth order gravity is bad defined since these models provide
$\omega_{BD}=0$ and this  is in contrast with observations
\cite{olmo_1,olmo_2}.

\begin{table}
\begin{center}
\begin{tabular}{|ccccc|} \hline
 ${\cal L}_{ST}$  & $\longrightarrow$ & $E+\phi$ & $\longleftarrow$ &
     ${\cal L}_{f(R)}$ \\
  $\downarrow$ &  & $\downarrow$ &  & $\downarrow$ \\
  ST -Eqs. & $\longrightarrow$ & Einstein Eqs. & $\longleftarrow$ & $f(R)$-Eqs. \\
  $\downarrow$ &  & $\downarrow$ &  & $\downarrow$ \\
  J-frame Sol. & $\longrightarrow$ & E-frame Sol. & $\longleftarrow$ &
  J-frame Sol. \\ \hline
\end{tabular}
\end{center}
\caption{Summary of the three approaches: Scalar-Tensor $(ST)$,
Einstein $+\varphi$, $(E+\phi)$ and $f(R)$  and their relations at
Lagrangians, field equations and solutions levels. The solutions
are in the Einstein frame for the minimally coupled case while
they are in Jordan frame for $f(R)$ and $ST$-gravity. Clearly,
$f(R)$ and $ST$ theories can be rigorously compared only recasting
them in the Einstein frame.}
\end{table}

Scalar-tensor theories and $f(R)$ theories can  be rigorously
compared, after conformal transformations, in the Einstein frame
where both  kinetic and potential terms are present. With this
consideration in mind, $F(\phi)$ and $f'(R)$ can be considered
analogous quantities in Jordan frame and then the PPN limit can be
developed \footnote{To be precise, conformal transformations
should be operated "before" performing PPN-limit and results
discussed in the same frame. A back conformal transformation,
after PPN limit, could be misleading due to gauge trobles.}.

Starting from this analogy,   the PPN results for scalar-tensor
gravity can be extended to  fourth order gravity \cite{ppn-noi}.
In fact, identifying $\phi\,\rightarrow\,R$ \cite{wands:cqg94}, it
is possible to extend the definition of the scalar-tensor
PPN-parameters \cite{damour1,schimd} to the case of fourth order
gravity\,:
\begin{equation}\label{ppn-R1}
\gamma-1=-\frac{{f''(R)}^2}{f'(R)+2{f''(R)}^2}\,, \qquad
\beta-1=\frac{1}{4}\left(\frac{f'(R)\cdot
f''(R)}{2f'(R)+3{f''(R)}^2}\right)\frac{d\gamma}{dR}.
\end{equation}
 In  \cite{ppn-noi}, these definitions have
been confronted with the observational upper limits on $\gamma$
and $\beta$ coming from Mercury Perihelion Shift \cite{mercury}
and Very Long Baseline Interferometry \cite{VLBI}. Actually, it is
possible to show that data and  theoretical predictions from Eqs.
(\ref{ppn-R1}) agree in the limits of experimental measures for
several classes of fourth order theories. Such a result tells us
that extended theories of gravity are not ruled out from Solar
System experiments but a more careful analysis of theories against
 experimental limits has to be performed. A possible procedure
could be to link the analytic form of a generic fourth order
theory with experimental data. In fact, the matching between data
and theoretical predictions, found in \cite{ppn-noi}, holds
provided some restrictions for the model parameters but gives no
general constraints on the theory. In general, the function $f(R)$
could contain an infinite number of parameters (i.e. it can be
conceived as an infinite  power series \cite{schmidtNL}) while, on
the contrary, the number of useful relations  is finite (in our
case we have only two relations). An attempt to deduce the form of
the gravity Lagrangian can be to consider the relations
(\ref{ppn-R1}) as differential equations for $f(R)$, so that,
taking into account the experimental results, one could constrain,
in principle, the model parameters by the measured values of
$\gamma$ and $\beta$.

%\section{Fourth Order Theories compatible with experimental limits on $\gamma$ and $\beta$}

\noindent 3. The  idea is  supposing the relations  for $\gamma$
and $\beta$ as differential equations. This hypothesis is
reasonable if the derivatives of $f(R)$ function are smoothly
evolving with the Ricci scalar. Formally, one can consider the
r.h.s. of the definitions (\ref{ppn-R1}) as differential relations
which have to be matched with values of PPN-parameters. In other
words, one has to solve the equations (\ref{ppn-R1}) where
$\gamma$ and $\beta$ are two parameters.  Based on such an
assumption, on can try to derive the  largest class of $f(R)$
theories compatible with experimental data. In fact, by the
integration of Eqs. (\ref{ppn-R1}), one obtains a solution
parameterized  by $\beta$ and $\gamma$ which have to be confronted
with the experimental quantities $\beta_{exp}$ and $\gamma_{exp}$.

Assuming  $f'(R)+2{f''(R)}^2\neq 0$ and defining ${\displaystyle
A\,=\,\Bigl|\frac{1-\gamma}{2\gamma-1}\Bigl|},$ we obtain from
(\ref{ppn-R1}) a differential equation for $f(R)$:
\begin{equation}
[f''(R)]^2=Af'(R)\,.
\end{equation}
The general solution of such an equation is a third order
polynomial $f(R)=aR^3+bR^2+cR+d$ whose coefficients have to
satisfy the conditions\,: $a=b=c=0$ and $d \neq 0$ (trivial
solution) or $a=\frac{A}{12}, b=\pm\frac{\sqrt{Ac}}{2}$, with $c,
d \neq 0$. Thus, the general solution for the non-trivial case, in
natural units, reads
\begin{equation}\label{ger-sol1}
f(R)=\frac{1}{12}\Bigl|\frac{1-\gamma}{2\gamma-1}\Bigl|R^3\pm\frac{\sqrt{c}}{2}
\sqrt{\Bigl|\frac{1-\gamma}{2\gamma-1}\Bigl|}R^2+cR+d\,.
\end{equation}
It is evident that the integration constants $c$ and $d$ have to be
compatible with GR prescriptions and, eventually, with the presence
of a cosmological constant. Indeed, when $\gamma \rightarrow 1$,
which implies $f(R) \rightarrow cR+d$,  the GR\,-\,limit is
recovered. As a consequence the values of these constants remain
fixed ($c$\,=\,1 and $d$\,=\,$\Lambda$, where $\Lambda$ is the
cosmological constant). Therefore, the fourth order theory provided
by Eq.(\ref{ger-sol1}) becomes
\begin{equation}\label{ger-sol2}
f_{\pm}(R)=
\frac{1}{12}\Bigl|\frac{1-\gamma}{2\gamma-1}\Bigl|R^3\pm
\frac{1}{2}\sqrt{\Bigl|\frac{1-\gamma}{2\gamma-1}\Bigl|}R^2+R+\Lambda\,,
\end{equation}
where we have formally displayed the two branch form of the
solution depending on the sign of the coefficient entering the
second order term. Since the constants $a, b, c, d$ of the general
solution satisfy the relation $3ac-b^2=0$, one can easily verify
that it gives\,:
\begin{equation}
\frac{d\gamma}{dR}\Bigl|_{f_{\pm}(R)}=-\frac{d}{dR}\frac{{f''(R)}^2}{f'(R)+2{f''(R)}^2}\Bigl|_{f_{\pm}(R)}\,=\,0\,,
\end{equation}
where the subscript $_{f_{\pm}(R)}$ refers the calculation to the
solution (\ref{ger-sol2}). This result, compared with the second
differential equation  Eq.(\ref{ppn-R1}), implies
$
4(\beta-1)=0\,,
$
which means the compatibility of the solution even with this
second relation.

%\section{Comparing with experimental measurements}

\noindent 4. Up to now we have discussed a family of fourth order
theories (\ref{ger-sol2}) parameterized by the PPN-quantity
$\gamma$; on the other hand, for this class of Lagrangians, the
parameter $\beta$ is compatible with  GR value being unity.

\begin{table}[ht]
\centering
\begin{tabular}{|l|c|}
\hline\hline
  Mercury Perihelion Shift& $|2\gamma-\beta-1|<3\times10^{-3}$ \\\hline
 Lunar Laser Ranging &  $4\beta-\gamma-3\,=\,-(0.7\pm 1)\times{10^{-3}}$ \\\hline
 Very Long Baseline Interf. &  $|\gamma -1|\,=\,4\times10^{-4}$ \\\hline
 Cassini Spacecraft &  $\gamma-1\,=\,(2.1\pm 2.3)\times10^{-5}$ \\\hline\hline
\end{tabular}
\caption{\small \label{ppn} A schematic resume of recent
experimental constraints on the PPN-parameters. They are the
perihelion shift of Mercury \cite{mercury}, the Lunar Laser
Ranging \cite{lls}, the upper limit coming from the Very Long
Baseline Interferometry \cite{VLBI} and the results obtained by
the estimate of the Cassini spacecraft delay into the radio waves
transmission near the Solar conjunction \cite{cassini}.}
\end{table}

\begin{figure}[htbp]
\centering
  \includegraphics[width=8cm]{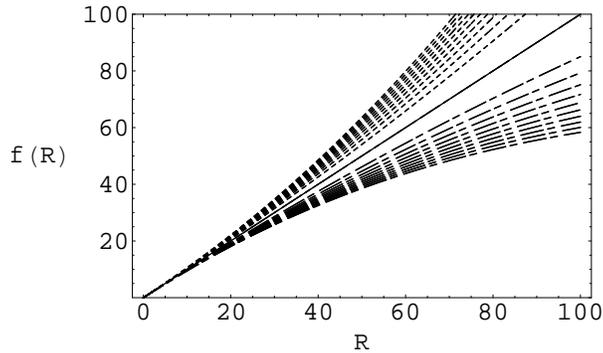}\\
  \caption{Plot of the two branch solution provided in Eq.(\ref{ger-sol2}). The $f_{+}(R)$ (dotted line) branch
  family is up to GR solution (straight line), while the one indicated with $f_{-}(R)$ (dotted-dashed line) remains below this line.
  The different plots for each family refer to different values of $\gamma$ fulfilling the condition $|\gamma-1|\leq 10^{-4}$
  and increased by step of $10^{-5}$.}
\label{fig1}
\end{figure}

Now, the further step  directly characterizes such a class of
theories by means of the experimental estimates of $\gamma$. In
particular, by fixing $\gamma$ to its observational estimate
$\gamma_{exp}$, we will obtain the weight of the coefficients
relative to each of the non-linear terms in the Ricci scalar of
the Lagrangian (\ref{ger-sol2}). In such a way, since GR
predictions require exactly $\gamma_{exp}\,=\,\beta_{exp}\,=\,1$,
in the case of fourth order gravity, one could to take into
account  small deviations from this values as inferred from
experiments. Some plots can contribute to the discussion of this
argument. In Fig.\ref{fig1},    the Lagrangian  (\ref{ger-sol2})
is plotted. It is parameterized for several values of $\gamma$
compatible with the experimental bounds coming from the Mercury
perihelion shift (see Table.1 and \cite{mercury}). The function is
plotted in the range $R\geq 0$. Since the property
$f_{+}(R)=-f_{-}(-R)$ holds for the function (\ref{ger-sol2}), one
can easily recover the shape of the plot in the negative region.
As it is reasonable, the deviation from GR becomes remarkable when
scalar curvature is large.

\begin{figure}[htbp]
\centering
  \includegraphics[width=8cm]{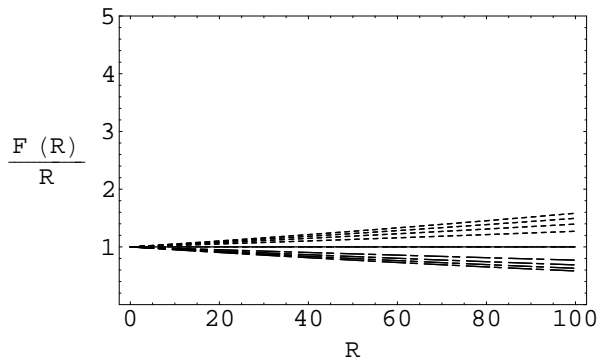}\\
  \caption{The ratio $f(R)/R$. It is shown the deviation of the fourth order gravity from GR considering the PPN-limit.
  Dotted and dotted-dashed lines refer to the $f_{+}(R)$ and $f_{-}(R)$ branches plotted with respect to several values
  of $\gamma$ (the step in this case is $2.5\times{10^{-5}}$). }
  \label{fig2}
\end{figure}

\begin{figure}[htbp]
\centering
  \includegraphics[scale=1]{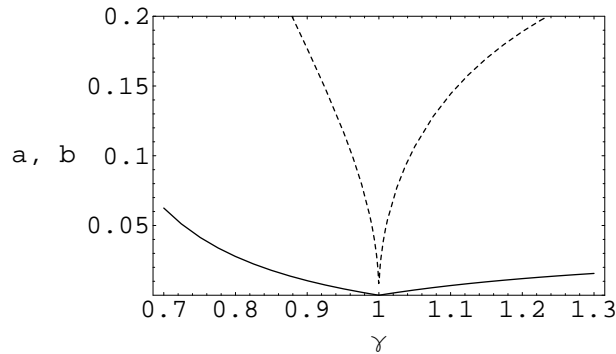}\\
  \caption{Plot of the modulus of coefficients: $a$ ($R^3$) (line) and $b$ ($R^2$) (dashed line). The choice
  of plotting the modulus of coefficients  is the consequence of two solution for $f(R)$.}
  \label{fig3}
\end{figure}

In order to display the differences between the theory
(\ref{ger-sol2}) and  Hilbert-Einstein one,   the ratio $f(R)/R$
is plotted in Fig.\ref{fig2}. Again it is evident that the two
Lagrangians differ significantly for great values of the curvature
scalar. It is worth noting that the formal difference between the
PPN-inspired Lagrangian and the GR expression can be related to
the physical meaning of the parameter $\gamma$ which is the
deviation from the  Schwarzschild-like solution. It measures the
spatial curvature of the region which one is investigating, then
the deviation  from the local flatness can be due to the influence
of higher order contributions in Ricci scalar.  On the other hand,
one can  reverse the argument and notice that if such a deviation
is measured, it can be recast in the framework of  fourth order
gravity, and in particular its ``amount" indicates the  deviation
 from GR. Furthermore, it is worth considering that, in the
 expression (\ref{ger-sol2}),
 the modulus  of the
 coefficients in $\gamma$ (i.e. the strength of the term)
 decreases by increasing the  degree of $R$. In
particular, the highest values of cubic and squared terms in $R$
are, respectively, of order $10^{-4}$ and $10^{-2}$ (see
Fig.\ref{fig3}) then GR remains a viable theory  at short
distances (i.e. Solar System) and low curvature regimes.

A remark is in order at this point. The class of theories which we
have discussed is a third order function  of the Ricci scalar $R$
parameterized by the experimental values of the PPN parameter
$\gamma$. In principle, any  analytic $f(R)$ can be compared with
the Lagrangian (\ref{ger-sol2}) provided suitable values of the
coefficients. However, more general results can be achieved
relaxing the condition $\beta=1$ which is an intrinsic feature for
(\ref{ger-sol2}) (see for example \cite{ppn-noi}). These
considerations suggest to take into account, as physical theories,
functions of the Ricci scalar which slightly deviates from GR,
i.e. $f(R)\,=\,f_0R^{(1+\epsilon)}$ with $\epsilon$ a small
parameter which indicates how much the theory deviates from GR
\cite{barrow}. In fact, supposing $\epsilon$ sufficiently small,
it is possible to approximate this expression
\begin{equation}\label{fReps} f_0|R|^{(1+\epsilon)}\simeq
f_0|R|\biggl(1+\epsilon\ln|R|+\frac{\epsilon^2\ln^2|R|}{2}+\dots\biggr)\,.
\end{equation}
This  relation can be easily confronted with the solution
(\ref{ger-sol2}) since, also in this case, the corrections have
very small ``strength" \cite{capfran}.

%\section{Conclusions}

\noindent 5. We have shown how a polynomial Lagrangian in the
Ricci scalar $R$, compatible with the PPN-limit, can be recovered
in the framework of fourth order gravity. The  approach is based
on the formulation of the PPN-limit of such gravity models
developed in analogy with scalar-tensor gravity \cite{ppn-noi}. In
particular,  considering the local relations defining the PPN
fourth order  parameters as differential expressions, one  obtains
a third-order polynomial in the Ricci scalar which is
parameterized by the PPN-quantity $\gamma$ and  compatible with
the limit $\beta\,=\,1$. The order of deviation from the linearity
in $R$ is induced by the deviations of $\gamma$ from the GR
expectation value $\gamma=1$. Actually, the PPN parameter $\gamma$
may represent the  key parameter  to discriminate among
relativistic theories of gravity. In particular, this quantity
should be significatively tested at Solar System scales by
forthcoming experiments like LATOR \cite{nordtvedt}. From a
physical point of view, any analytic function of $R$, by means of
its Taylor expansion, can be compared with (\ref{ger-sol2}).
Therefore, a theory like $f(R)\,=\,f_0R^{(1+\epsilon)}$,
indicating small deviations from standard GR, is in agreement with
the proposed approach, so, in principle, the experimental $\gamma$
could indicate the value of the parameter $\epsilon$. In
conclusion,  one can reasonably state
 that generic fourth-order  gravity models could be viable candidate theories
 even in the PPN-limit. In other words, due to the presented results, they cannot be  a priori
excluded  at Solar System scales.

\end{document}